\documentclass[a4paper,11pt]{article}
\usepackage{pos}

\title{Cosmological implications of EW vacuum instability: constraints on the Higgs-curvature coupling from inflation}
\ShortTitle{Vacuum decay $\xi$-constraints from inflation}

\author*[a]{Andreas Mantziris}

\affiliation[a]{Imperial College London,\\
  180 Queen's Gate, London, United Kingdom}

\emailAdd{a.mantziris18@imperial.ac.uk}

\abstract{The current experimentally measured parameters of the Standard Model (SM) suggest that our Universe lies in a metastable electroweak vacuum, where the Higgs field is prone to vacuum decay to a lower state with catastrophic consequences. Our measurements dictate that such an event has not taken place yet, despite the many different mechanisms that could have triggered it in our past light-cone. The focus of our work has been to calculate the probability of the false vacuum to decay during the period of inflation and use it to constrain the last unknown renormalisable SM parameter $\xi$, which couples the Higgs field with space-time curvature. More specifically, we derived lower $\xi$-bounds from vacuum stability in three inflationary models: quadratic and quartic chaotic inflation, and Starobinsky-like power-law inflation. We also took the time-dependence of the Hubble rate into account both in the geometry of our past light-cone and in the Higgs effective potential, which is approximated with three-loop renormalisation group improvement supplemented with one-loop curvature corrections.}

\FullConference{%
  *** The European Physical Society Conference on High Energy Physics (EPS-HEP2021), ***\\
  *** 26-30 July 2021 ***\\
  *** Online conference, jointly organized by Universität Hamburg and the research center DESY ***
}

\def\Nbub{\langle {\cal N}\rangle}

\begin{document}
\maketitle

\section{Introduction}
In 2012, the last missing particle of the Standard Model (SM) of particle physics, the Higgs boson, was observed in the Large Hadron Collider at CERN, henceforth establishing firmly the validity of the SM as a self-consistent theory of fundamental particles and their interactions. The SM has famously provided predictions about observables, which agree with experimental results to many decimal places and therefore render it as the most successful physical theory so far. Moreover, the Higgs boson has a number of interesting characteristics that may have allowed it to affect the evolution of the Universe \cite{Markkanen:2018pdo}. In the context of the SM, the eponymous boson is the excitation of the Higgs field that permeates space-time and via its coupling to matter fields, it generates their masses. This is famously known as the Higgs mechanism and it results from the spontaneous symmetry breaking of the $SU(2) \times U(1)$ symmetry of the electroweak (EW) force into electromagnetism's $U(1)$ symmetry, due to the Higgs' non-zero vacuum expectation value (vev) of $v \approx 246$ GeV. 

\begin{figure}[h]
\centering
\includegraphics[width=0.87\linewidth]{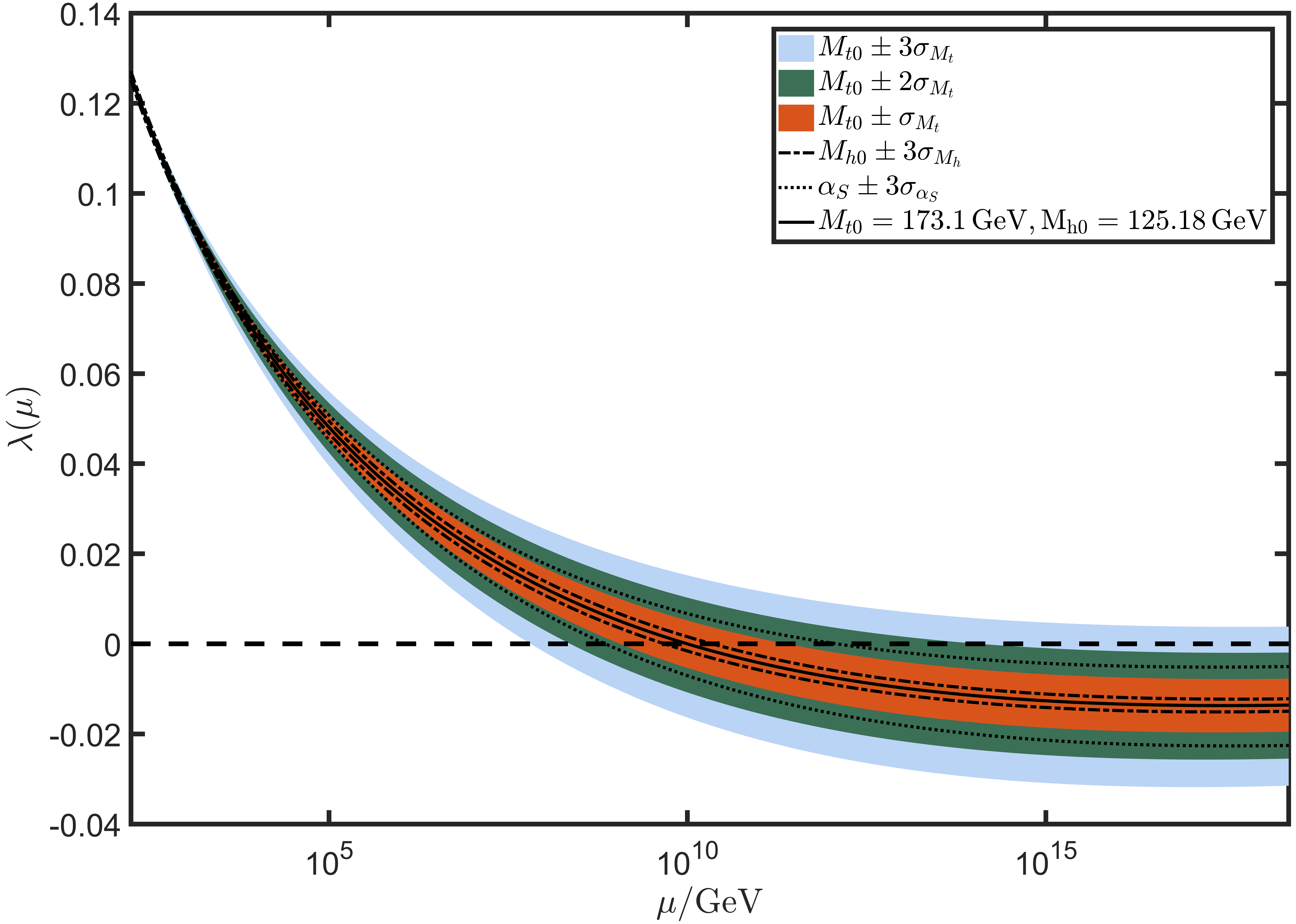}
\caption{Running of the Higgs self-interaction coupling, where the shaded regions correspond to $3 \sigma$ variance due to the uncertainty in the measurement of the top quark mass $m_t$. \cite{Markkanen:2018pdo} }
\label{Fig1}
\end{figure}

The experimentally measured masses of the SM particles lie in a range, where the Higgs self-interaction $\lambda$ does not diverge before the Planck scale, as it runs with the energy scale $\mu$ \cite{Buttazzo:2013uya}.  As we can see in figure \ref{Fig1}, the four-point coupling is a smooth function, without any poles or discontinuities all the way up to the high energy scales beyond $10^{15}$ GeV. This implies that the SM might be a consistent minimal model that could describe our very early Universe until quantum gravity effects become significant \cite{Ellis:2009tp}. The running of $\lambda$ depends on the SM particles' masses, however the Higgs and the top quark, being the heaviest of each kind, dominate their respective contribution.

\begin{figure}[h]
\centering
\includegraphics[width=0.55\linewidth]{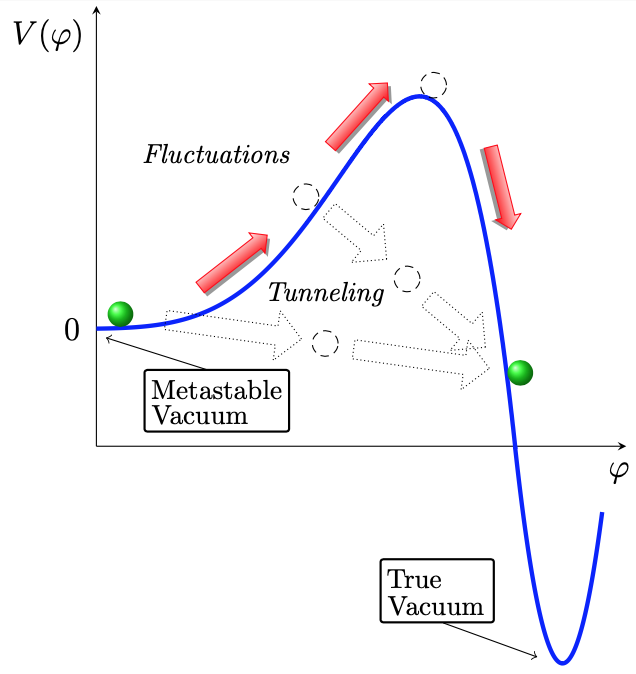}
\caption{Vacuum decay for a double well potential of a scalar field $\varphi$, from a metastable vacuum to its true vacuum state via quantum tunnelling, thermal fluctuations, or a combination of both. \cite{Markkanen:2018pdo}}
\label{Fig2}
\end{figure}

Furthermore, for most of the curves shown in figure \ref{Fig1}, where the coloured bands emphasize the uncertainty in the measurement of $m_t$, it is evident that the self-interaction switches sign beyond approximately $10^{10}$ GeV. This means that the famous quartic Higgs potential
\begin{align}
	V_H (h, \mu) =  \frac{\lambda(\mu)}{4} h^4
\end{align}
will ``turn'' and develop a second, lower vacuum state, separated by a potential barrier, as depicted in figure \ref{Fig2}. Because the Higgs field is a quantum field, there is always a finite and non-zero probability for the field to tunnel through the barrier, on top of the classical fluctuations that could excite it above it. This renders the current EW vacuum state, that the Higgs resides in, a metastable one, which has survived throughout our cosmological history, despite the many different mechanisms that could have triggered its decay. However, it is important to note that we cannot rule out the absolute stability of the EW vacuum yet, as the uncertainty in the measurement of $m_t$ is still significant.

\section{Vacuum decay in the early universe}
The problem with a lower ground state at high field values, lies in the fact that our Universe would, at the very least, look very different because the masses of all particles would be different. In addition, the violent process of the decay itself, induces the nucleation of bubbles of true vacuum, which for all practical purposes within the SM are assumed to contain a singularity \cite{Espinosa:2015qea}. Therefore, as they rapidly expand after formation with velocity close to the speed of light, they devour everything in their path without warning. A vacuum decay event would have catastrophic consequences for our false vacuum Universe and hence, the observation of its metastability allows us to infer something about fundamental physics. This follows from the argument that there could not have been any bubbles in our past light-cone, otherwise they would have already devoured us and the Higgs vev would not be at the EW scale. This means that the probability for the formation of zero bubbles has to be significant enough,  $P({\cal N} = 0) \sim {\cal O} (1) $. Because these nucleation events are rare, their probability follows a Poisson distribution, which is proportional to $e^{- \Nbub}$, where $\Nbub$ is the expectation value of the number of true vacuum bubbles. Therefore, there is an observational requirement  on their expectation value that $ \Nbub \leq 1$, which allows us to place constraints on fundamental physics.

The aforementioned expectation value is a product of the decay rate $\Gamma$ and the space-time volume element $d \mathcal{V}$, integrated over our past light-cone
 \begin{align}
    d\Nbub = \Gamma d{\cal V} \, \Rightarrow \, \, \Nbub = \int_{\substack{ \rm past }} d^4x \sqrt{-g} \Gamma(x) \, .
    \label{eq:Nbub-general}
 \end{align}
Thankfully, the decay rate is so low today, that the lifetime of our metastable vacuum is longer than the age of the Universe \cite{Rajantie:2016hkj}. However, there could have been significantly higher rates in the early Universe, and thus we are motivated to study vacuum decay during the period of cosmological inflation. With the seminal work of \cite{Callan:1977pt} and \cite{Coleman:1980aw} among others, classical solutions to the decay process from false to true vacuum were found. They are called instantons and in our case, where the inflationary Hubble rates are high, the Coleman-de Lucia instanton approaches the much simpler Hawking-Moss (HM) instanton \cite{Hawking:1981fz} with action difference and  decay rate given respectively by
        \begin{align}
            B_{\rm HM}(R) \approx \frac{384 \pi^2  \Delta V_{\rm H} }{R^2} \, , \quad              \Gamma_{\rm HM}(R) \approx \left(\frac{R}{12}\right)^2 e^{- B_{\rm HM}(R)} \, ,
            \label{eq:Gamma}
        \end{align}        
where $R$ is the Ricci scalar and $ \Delta V_{\rm H} = V_{\rm H}(h_{\rm bar}) - V_{\rm H}(h_{\rm fv})$ is the height of the barrier separating true from false vacuum (fv) in the Higgs potential $V_H$.

For this inflationary context, it is useful to express the integral (\ref{eq:Nbub-general}) in terms of the number of $e$-foldings of inflation $N = \mathrm{ln} \left(a_{\rm inf} / a(\eta) \right)$, where $a_{\rm inf}$ is the scale factor at the end of inflation, as
\begin{align}
    \Nbub =  \frac{4\pi}{3}\int_0^{N_{\mathrm{start}}} dN \left( \frac{a_{\mathrm{inf}} \left(\eta_0-\eta\left(N\right)\right)}{e^{N}} \right)^3 \frac{\Gamma(N)}{H(N) } \, ,
    \label{eq:Nbub}
\end{align}
where $\eta_0$ is the conformal time today, and $H$ is the Hubble rate. We are integrating backwards in time, from the end of inflation set at $N_{\rm inf} = 0$ towards its beginning at $N_{\rm start}$, for a duration of a least 60 $e$-foldings according to observations. By imposing the requirement that $\Nbub \leq 1$, with the minimal assumption that  $N_{\rm start} = 60$, we can constrain the remaining free parameter in (\ref{eq:Nbub}). 
 
In this case, it is the non-minimal coupling $\xi$, which couples the Higgs field $h$ with space-time curvature $R$ in the Higgs potential, and it enters the calculation via the potential as described in section \ref{sec:3}. The significance of constraining $\xi$ comes from the fact that it is the last unknown renormalisable parameter of the SM, as it cannot be probed with accelerator experiments, because space-time is not curved enough. On the other hand, space-time was significantly more curved in the early Universe and thus, studying the metastability of the EW vacuum during inflation can provide constraints that are orders of magnitude stronger than other methods \cite{Mantziris:2020rzh}.

\section{The effective Higgs potential in curved space-time} \label{sec:3}
It is necessary to embed the SM in a curved background as we are studying the early Universe. Therefore, besides the characteristic quartic term of the Higgs potential, curvature effects also enter at tree-level via the non-minimal coupling $\xi$ \cite{Chernikov:1968zm}, resulting in the potential given by
 \begin{align}
           V_{\rm H}(h, \mu, R) =  \frac{\xi(\mu)}{2}R h ^2 + \frac{\lambda(\mu)}{4} h^4 \, ,
        \end{align}
where we have made explicit the scale dependence of the couplings. Beyond tree-level, curvature effects enter via the loop corrections as well. If we include Minkowski terms to 3-loops and curvature corrections in de Sitter (dS) space at 1-loop, the potential reads
        \begin{align}
            V_{\rm H}(h, \mu, R) = \frac{\xi (\mu) }{2}Rh^2 + \frac{\lambda (\mu) }{4}h^4 + \frac{\alpha (\mu) }{144} R^2 + \Delta V_{\rm loops} (h, \mu, R) \, ,
        \end{align}
where the $\alpha$-term contains purely gravitational terms calculated in dS, that are radiatively generated in curved space proper renormalisation. The loop contribution is an extended sum over all degrees of freedom of the SM
    \begin{align}
     \Delta V_{\rm loops} = \frac{1}{64\pi^2} \sum\limits_{i=1}^{31}\bigg\{ n_i\mathcal{M}_i^4 \bigg[\log\left(\frac{|\mathcal{M}_i^2 |}{\mu^2}\right) - d_i \bigg] +\frac{n'_i R^2}{144}\log\left(\frac{|\mathcal{M}_i^2 |}{\mu^2}\right)\bigg\}  \, ,
\end{align}
where $\mathcal{M}_{i}$ refers to each effective mass with its curvature correction, as presented in detail in \cite{Markkanen:2018bfx}.

We can simplify the potential by removing the ``unphysical'' $\mu$-dependence, using Renormalisation Group Improvement (RGI). This method provides an approximation for the scale choice, by choosing the RG scale  $\mu=\mu_*(h,R)$ so that the loop correction vanishes, i.e. $\Delta V_{\rm loops} (h, \mu_*, R) = 0 $. This is a numerically heavy calculation, where we have to take into account the entirety of the SM particle spectrum and the running of all couplings, with the accompanying pole-matching \cite{Bezrukov:2009db} and $\beta$-functions \cite{Chetyrkin:2012rz}, resulting finally in a function for $\mu_*$ that cannot be written analytically. In the end, we have managed to obtain the state-of-the-art RG improved effective Higgs potential given by
\begin{align}
    V_{\rm H}^{\rm RGI}(h, R) = \frac{\xi(\mu_*(h, R))}{2}  R h ^2 +\frac{\lambda(\mu_*(h, R))}{4}  h^4 + \frac{\alpha(\mu_*(h,R))}{144} R^2 \, .
    \label{eq:VRGI}
\end{align}

\section{Results}
Ultimately, we have all the ingredients to complete our calculation that provides lower bounds on the Higgs curvature coupling. First, we evaluate the barrier height $ \Delta V_{\rm H} $ using (\ref{eq:VRGI}), in order to compute the decay rate (\ref{eq:Gamma}). Afterwards, given an inflationary model with potential $V(\phi)$ for the inflaton field, we can calculate, without any slow-roll approximation, the cosmological quantities contained in the integral (\ref{eq:Nbub}), i.e. the comoving radius of the past light-cone $\eta_0 - \eta(N)$, $H(N)$ and $a_{\rm inf}$. Finally, by imposing $\Nbub \leq 1$ on the integral, we can constrain $\xi$ from below as $\xi \geq \xi_{\Nbub=1}$.

\begin{figure}[h]
\centering
\includegraphics[width=0.8\linewidth]{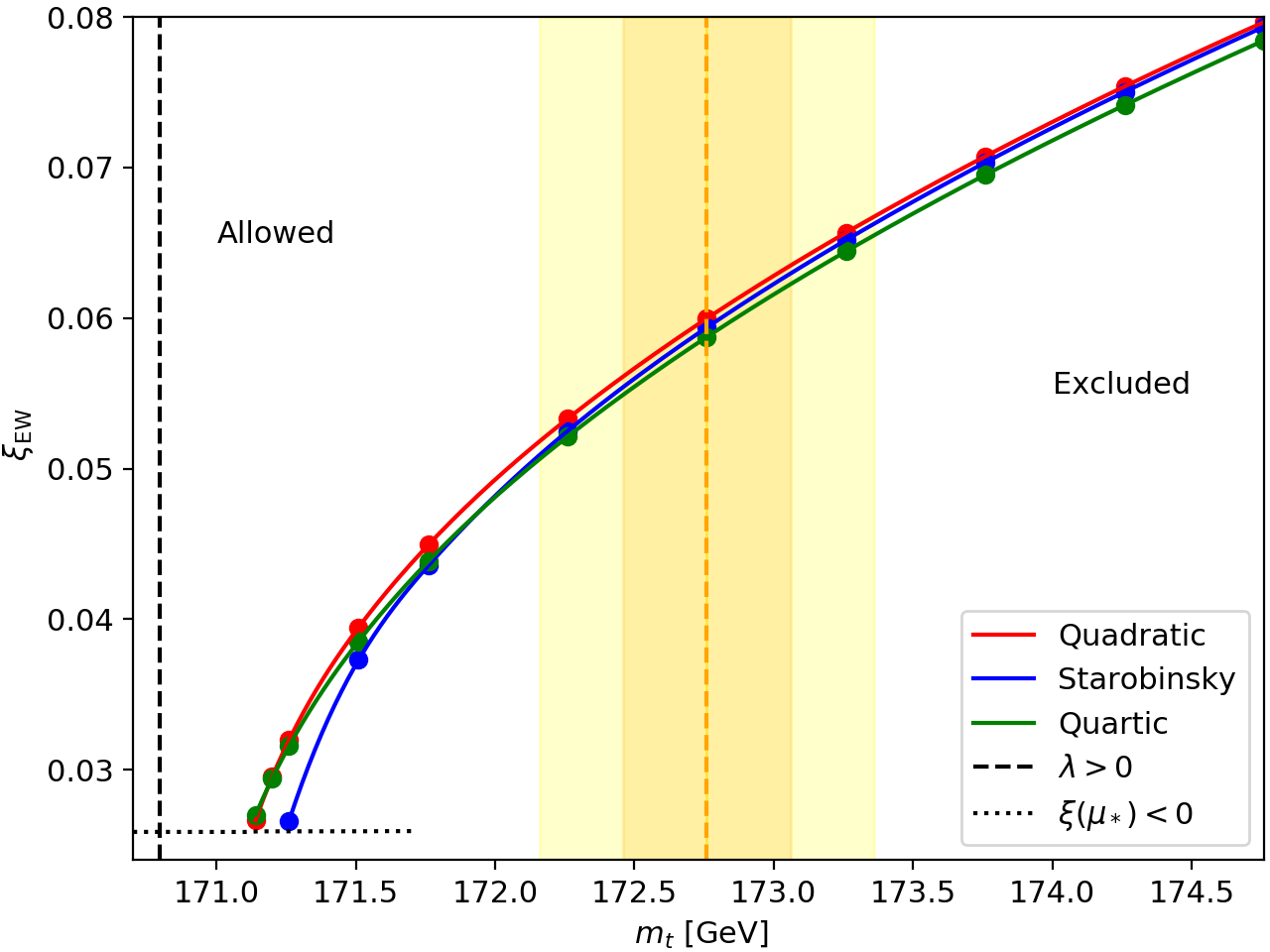}
\caption{Bounds on the curvature coupling at the EW scale with varying top quark mass for three inflationary models. The vertical orange line and its bands stand at the central value $m_t = (172.76 \pm 0.30)$ GeV, with $\sigma$ and $2 \sigma$ variance. The vertical black line lies at the threshold below which there is no lower ground state in the potential. The horizontal line signifies the value below which $\xi(\mu)$ turns negative as it runs. \cite{Mantziris:2020rzh}}
\label{fig:3}
\end{figure}

The lower bounds for quadratic inflation $V(\phi) = \frac{1}{2}m^2\phi^2$, quartic inflation $V(\phi) = \frac{1}{4} \lambda \phi^4 $, and Starobinsky-like \cite{Starobinsky:1979ty} power-law inflation $V(\phi) = \frac{3}{4} \alpha^2 M_P^4 \left( 1 - e^{-\sqrt{\frac{2}{3}}\frac{\phi}{M_P}} \right)^2$ are shown in figure \ref{fig:3}. Because the non-minimal coupling is a running parameter, we have bounded its value at the EW scale $\xi_{\rm EW} = \xi(\mu_{\rm EW})$, while also showing the effect of the uncertainty in the top quark mass $m_t$. It is important to note that the range of validity of our approach is limited by two factors, which are illustrated by the black horizontal and vertical lines at the lower left corner of figure \ref{fig:3}. Firstly, the vertical, dashed line denotes the lightest the top quark can be, that still results in a self-coupling $\lambda$ that turns negative as it runs, which therefore leads to the development of a second vacuum state. Hence, for smaller masses, the EW vacuum is absolutely stable and therefore it is not possible to use this analysis that assumes metastability. On the other hand, the vertical, dotted line depicts the minimum value for which $\xi(\mu)$ will remain positive as it runs. Below this threshold, the first term of (\ref{eq:VRGI}) switches sign during its run, with the vacuum being potentially destabilised. As our analysis did not account for this possibility, the curves terminate at this line. 

\section{Conclusions}
In conclusion, we have presented a method for using a state-of-the-art RG improved effective Higgs potential on curved space, with leading time-dependent curvature corrections for all SM constituents, that consistently includes Minkowski terms to 3-loops and 1-loop curvature corrections beyond dS, in order to obtain the most accurate constraints on the Higgs curvature coupling to date. These are summarised for all inflationary models considered as $\xi_{\rm EW} \gtrsim 0.06$, where their mild dependence on the top quark mass is evident from figure \ref{fig:3}. Moreover, the different inflationary models led to approximately degenerate results, due to their tuning according to the CMB measurements and the use of the dS approximation at certain steps of the calculation. Finally,  as properly explained in \cite{Mantziris:2020rzh}, these results are independent of the total duration of inflation beyond 60 $e$-foldings, with predominant bubble production taking place close to the end of inflation.

\acknowledgments
The author would like to thank Arttu Rajantie and Tommi Markkanen for the supervision, guidance and collaboration that led to the publication of \cite{Mantziris:2020rzh}, which served as the basis for the parallel talk presented at the European Physical Society Conference on High Energy Physics 2021. AM was supported by an STFC PhD studentship.

\bibliographystyle{JHEP}
\bibliography{references.bib}

\end{document}